\documentclass[letter,twocolumn,dvipdfmx]{jpsj3}

\usepackage[dvipdfmx]{graphicx}
\usepackage{amsmath,amssymb,bm}
\usepackage{color}

\usepackage{ulem}

\makeatletter
\newcommand\newblock{\hskip .11em\@plus.33em\@minus.07em}
\makeatother

\title{Atomic Charge Chirality from Electric Multipoles}

\author{Satoru Hayami and Rikuto Oiwa}
\inst{Department of Physics, Hokkaido University, Sapporo 060-0810, Japan}

\abst{
We establish an electric-multipole ($Q_{lm}$) criterion for chirality of local atomic charge distributions.
For the 11 crystallographic chiral point groups, we identify representative lowest-rank multipole contents required to distinguish the opposite handednesses.
For the axial groups $C_n$ and $D_n$ with $n=2,3,4,6$, chirality arises from the relative orientation of two azimuthal anisotropies with $m=n$, belonging to consecutive ranks $l=n$ and $n+1$. 
Their coexistence produces a latitude-dependent twist of the $n$-fold charge distribution.
The polar component, although allowed in $C_n$, is distinct from this chiral mechanism and is not required for handedness.
The cubic groups $T$ and $O$ exhibit distinct realizations, with octahedral chirality first appearing at rank nine.
We further show that, within a spinless $s$-$p$-$d$-$f$ atomic basis, the scalar charge density contains multipoles only up to rank six and therefore cannot encode the handedness of $C_6$, $D_6$, or $O$.
These results connect electric-multipole content and atomic orbital degrees of freedom to the microscopic encoding of crystallographic handedness.
}

\begin{document}
\maketitle

Chirality, which characterizes the distinction between an object and its mirror image, is a fundamental concept in physics, chemistry, and materials science~\cite{kelvin1894molecular, Barron1986jacs_asymmetric_systhesis, Barron_2004, kelvin2010baltimore, bousquet2025structural}.
In condensed matter systems, chirality gives rise to various phenomena, including electrical magnetochiral effect~\cite{Rikken_PhysRevLett.87.236602, yokouchi2017electrical, Aoki_PhysRevLett.122.057206, rikken2022dielectric}, optical activity~\cite{lipkin1964existence, kibble1965conservation, proskurin2017optical, mun2020electromagnetic, hayami2024analysis, Kanda_PhysRevB.111.035130}, magnetoelectric responses~\cite{yoda2015current, furukawa2017observation, Furukawa_PhysRevResearch.3.023111}, and conversion between time-reversal-even polar and axial quantities~\cite{Oiwa_PhysRevLett.129.116401, hayami2025chirality}.
In magnetism, chiral crystal structures can host characteristic noncollinear spin textures~\cite{Muhlbauer_2009skyrmion, yu2010real, nagaosa2013topological, Tokura_doi:10.1021/acs.chemrev.0c00297, hayami2024stabilization}.
For example, magnetic skyrmions have been observed in chiral magnets such as MnSi~\cite{ishikawa1976helical,Muhlbauer_2009skyrmion,Neubauer_PhysRevLett.102.186602,Jonietz_skyrmion,Adams_PhysRevLett.107.217206},  Fe$_{1-x}$Co$_x$Si~\cite{yu2010real,beille1983long}, and the $f$-electron compound EuPtSi~\cite{kakihana2018giant,kaneko2019unique,tabata2019magnetic,kakihana2019unique,hayami2021field,Matsumura_PhysRevB.109.174437}.

Beyond chirality originating directly from the crystal structure, recent attention has focused on electronic chirality, in which handedness emerges from electronic degrees of freedom~\cite{Oiwa_PhysRevLett.129.116401, yan2024structural, qin2026molecular, winkler2026standard}.
Such electronic chiral states can arise spontaneously even when the high-temperature crystal structure itself is achiral.
More broadly, electronically driven odd-parity multipolar order provides a route to the emergence of chirality. 
For example, the spin-orbit-coupled metal Cd$_2$Re$_2$O$_7$ exhibits an electronically driven odd-parity multipolar state accompanied by inversion-symmetry breaking~\cite{yamaura2002low, Matteo_PhysRevB.96.115156, hiroi2018pyrochlore, Hayami_PhysRevLett.122.147602, hirai2022successive}, while recent NMR measurements on URhSn have revealed a spontaneous electronic chiral order associated with antiferroic quadrupolar ordering~\cite{harima2023hidden, tabata2025successive, ishitobi2026purely}.
Local electronic chirality has also been discussed in connection with the hidden-order phase of URu$_2$Si$_2$, where the breaking of local reflection symmetries and a chiral charge order have been proposed~\cite{Kambe_PhysRevB.97.235142, kambe2020symmetry, hayami2023chiral}.
More generally, recent developments have established microscopic approaches to characterizing pseudoscalar electronic degrees of freedom at both the single-particle level~\cite{Oiwa_PhysRevLett.129.116401, Miki_PhysRevLett.134.226401} and in interacting many-body systems~\cite{miki2026electron, kuniyoshi2026theory}.
These developments motivate a microscopic characterization of chirality directly in terms of electronic degrees of freedom.

A complementary experimental development has made it possible to probe such electronic chirality directly in real space. 
Recent synchrotron X-ray diffraction measurements combined with the core differential Fourier synthesis method~\cite{kitou2017cdfs, kitou2020cdfs} have visualized strongly anisotropic and asymmetric valence charge distributions in the cubic chiral crystal CoSi~\cite{aoyagi2026unraveling, aoyagi2026experimental}.
The observed charge anisotropy reflects the underlying chiral bonding network and changes with crystallographic handedness.
In particular, the complex hybrid orbital decomposition approach introduced for B20-type $M$Si ($M={\rm Cr,Mn,Fe,Co}$) quantifies the local chiral twist through the relative phase of symmetry-allowed orbital-hybridization terms~\cite{aoyagi2026experimental}.
These observations raise a fundamental question: under what conditions can the handedness of a local crystallographic environment actually be encoded in the angular distribution of the electronic charge density around an atomic site?

In this Letter, we formulate an electric-multipole ($Q_{lm}$) criterion for chirality of local atomic charge distributions.
By systematically examining the 11 crystallographic chiral point groups, we identify representative lowest-rank combinations of electric multipoles required for a scalar charge density to distinguish the opposite handednesses. 
For the axial groups $C_n$ and $D_n$, we show that chirality arises from the relative orientation of azimuthal anisotropies with the same index $m=n$ and consecutive ranks $l = n
$ and $n+1$, producing a latitude-dependent twist of the charge distribution. 
We further distinguish this chiral coupling from the polar electric dipole $Q_{10}$ and clarify the distinct multipole realizations in the tetrahedral and octahedral chiral groups.
Finally, we show that a single-center spinless $s$-$p$-$d$-$f$ basis cannot encode the handedness of $C_6$, $D_6$, or $O$ in the scalar charge density because the required multipoles lie beyond rank six.
Our results establish a direct connection between crystallographic handedness, atomic orbital content, and local electronic charge chirality.

Let us start by considering a scalar charge density around a single center,
\begin{align}
 \rho(\bm r)
   =\rho_0(r)+\sum_{lm\gamma}q_{lm}^{(\gamma)}
  Q_{lm}^{(\gamma)}(\bm r),
   \label{eq:rho}
\end{align}
where $\rho_0(r)$ denotes the spherically symmetric part of the charge density, $l$ and $m$ are the rank and azimuthal index of the electric multipole, respectively, and $q_{lm}^{(\gamma)}$ is its amplitude.
Here, $l=1,2,\ldots$ and $m=0,1,\ldots,l$, while $\gamma=c,s$ labels the cosine- and sine-type real tesseral components for $m>0$, respectively.
Ignoring normalization and radial factors~\cite{Kusunose_JPSJ.77.064710,Hayami_PhysRevB.98.165110, Watanabe_PhysRevB.98.245129},
\begin{align}
 Q_{lm}^{(c)}
 &\propto r^lP_l^m(\cos\theta)\cos m\phi,\\
 Q_{lm}^{(s)}
 &\propto r^lP_l^m(\cos\theta)\sin m\phi , 
 \label{eq:tesseral}
\end{align}
where $P_l^m(\cos\theta)$ denotes the associated Legendre polynomial.
For $m = 0$, only one real component exists, which we denote by $Q_{l0}$.
For fixed $(l,m)$, the $c$ and $s$ components represent the same $m$-fold anisotropy with different azimuthal orientations and are mixed by rotations about the $z$ axis.
The electric multipole of rank $l$ has spatial parity $(-1)^l$.

Chirality, on the other hand, is characterized by a time-reversal-even pseudoscalar: it is invariant under proper rotations and changes sign under improper spatial operations.
Within the present multipole description, such a pseudoscalar must emerge from an appropriate combination of polar electric multipoles.
Combining even- and odd-rank electric multipoles provides the required inversion-odd character, but this alone does not guarantee chirality, since an improper symmetry operation may still remain.
The key is therefore to identify lowest-rank combinations of electric multipoles whose coexistence removes all improper symmetry operations and thereby gives rise to a pseudoscalar measure of chirality.

Among the 32 crystallographic point groups, the 11 chiral point groups containing only proper rotations are
\begin{align}
 C_1,C_2,C_3,C_4,C_6,D_2,D_3,D_4,D_6,T,O.
 \label{eq:chiralPG}
\end{align}
We first consider the axial groups $C_{n}$ and $D_{n}$ with a principal $n$-fold axis.  
Invariance under the principal $n$-fold rotation $C_{nz}$ restricts $m$ of $Q_{lm}^{(c/s)}$ to integer multiples of $n$.
Since $|m| \leq l$, only $m = 0$ is possible for $l < n$; the corresponding multipoles are axially symmetric and hence cannot produce a chiral charge distribution.
The lowest-rank azimuthally anisotropic components therefore appear at $l = n$~\cite{Hayami_PhysRevB.98.165110, Watanabe_PhysRevB.98.245129},
\begin{align}
 Q_{nn}^{(c)}&\propto \sin^n\theta\cos n\phi,\\
 Q_{nn}^{(s)}&\propto \sin^n\theta\sin n\phi.
 \label{eq:Qnn}
\end{align}
An arbitrary linear combination within this rank can always be expressed as
\begin{align}
 a_cQ_{nn}^{(c)}+a_sQ_{nn}^{(s)}
 \propto A\sin^n\theta\cos(n\phi-\delta_a),
 \label{eq:single}
\end{align}
which corresponds simply to an azimuthal rotation of the $n$-fold pattern and retains at least a vertical mirror symmetry. 
Here, $a_c$, $a_s$, and $A$ are numerical coefficients.  
Thus, the rank-$n$ sector alone cannot support chirality.

Since an additional anisotropic component with a different polar-angle dependence is therefore required, we next consider the lowest-rank symmetry-allowed sector with the same azimuthal index $m = n$, which occurs at rank $l = n+1$:
\begin{align}
 Q_{n+1,n}^{(c)}
 &\propto
 \sin^n\theta\cos\theta\cos n\phi,\\
 Q_{n+1,n}^{(s)}
 &\propto
 \sin^n\theta\cos\theta\sin n\phi.
 \label{eq:Qn1n}
\end{align}
The most general lowest-rank anisotropic charge density involving these two sectors is then
\begin{align}
 \rho_n =
 a_cQ_{nn}^{(c)}+a_sQ_{nn}^{(s)}
 +b_cQ_{n+1,n}^{(c)}+b_sQ_{n+1,n}^{(s)}.
 \label{eq:general}
\end{align}
Additional $m=0$ multipoles may also be symmetry allowed. 
However, because they are azimuthally symmetric, they do not contribute to the minimal handedness measure discussed below.
Introducing the coefficient vectors
\begin{align}
 \bm a=(a_c,a_s),\qquad
 \bm b=(b_c,b_s),
\end{align}
we define
\begin{align}
 \chi_n=\det(\bm a,\bm b)=a_cb_s-a_sb_c.
 \label{eq:chi}
\end{align}
A proper rotation about the principal axis leaves $\chi_n$ invariant, whereas a mirror reflection reverses its sign.
For a vertical mirror, this sign reversal originates from the transformation between the cosine- and sine-type components, whereas for a horizontal mirror it follows from the opposite parities of the rank-$n$ and rank-$(n+1)$ sectors.
Within the symmetry class constrained by the principal $C_n$ axis, a finite $\chi_n$ therefore excludes the improper operations compatible with the achiral supergroups of $C_n$ or $D_n$.
Consequently, $\chi_n$ serves as a time-reversal-even pseudoscalar handedness descriptor.

To clarify its geometrical meaning, we parametrize the two coefficient vectors as
\begin{align}
 \bm a&=A(\cos\delta_a,\sin\delta_a),\\
 \bm b&=B(\cos\delta_b,\sin\delta_b),
\end{align}
where $\delta_a$ and $\delta_b$ are the phase angles of the rank-$n$ and rank-$(n+1)$ anisotropies; their physical azimuthal orientations are $\delta_a/n$ and $\delta_b/n$, respectively.
Then,
\begin{align}
 \chi_n=AB\sin(\delta_b-\delta_a).
 \label{eq:phase}
\end{align}
Thus, $\chi_n$ measures the relative orientation of the two anisotropic sectors.

For $C_3$, this construction is directly connected to the chirality descriptor introduced for B20-type compounds~\cite{aoyagi2026experimental}.  
The orbital coherence $p_{m=1}^{*}d_{m=-2}$ carries $|m|=3$ and odd parity and therefore belongs to the rank-3 electric-multipole sector $Q_{33}^{(c/s)}$, whereas $d_{m=1}^{*}d_{m=-2}$ carries $|m|=3$ and even parity and belongs to the rank-4 sector $Q_{43}^{(c/s)}$.  
Thus, up to normalization and phase conventions, the B20 chirality descriptor has the same structure as $\chi_3$.

The geometrical meaning of $\chi_n$ becomes transparent by expressing the charge distribution in terms of its angular dependence.
Equation~(\ref{eq:general}) can be written as
\begin{align}
 \rho_n
 \propto
 \sin^n\theta
 \left(
 A_\theta\cos n\phi+B_\theta\sin n\phi
 \right),
 \label{eq:angular}
\end{align}
where $A_\theta=a_c+b_c\cos\theta$ and $B_\theta=a_s+b_s\cos\theta$.
Equivalently,
\begin{align}
 \rho_n
 \propto
 \sin^n\theta R(\theta)
 \cos[n\phi-\delta(\theta)],
 \label{eq:twist}
\end{align}
where $R(\theta)=\sqrt{A_\theta^2+B_\theta^2}$ and
\begin{align}
 \tan\delta(\theta)
 =
 \frac{a_s+b_s\cos\theta}
      {a_c+b_c\cos\theta}.
 \label{eq:delta}
\end{align}
The angle $\delta(\theta)/n$ specifies the azimuthal orientation of the $n$-fold anisotropic pattern at a given polar angle $\theta$.
Introducing $x=\cos\theta$, its variation is given by
\begin{align}
 \frac{d}{dx}\left(\frac{\delta}{n}\right) = \frac{\chi_n}{n\left[(a_c+b_cx)^2+(a_s+b_sx)^2\right]}.
 \label{eq:twistrate}
\end{align}
Thus, a finite $\chi_n$ produces a latitude-dependent rotation of the $n$-fold anisotropy, with its sign determining the sense of the twist.  
When $\chi_n=0$, the orientation is independent of $\theta$, leaving a common vertical mirror plane.
Therefore, within this minimal sector, $\chi_n$ provides a signed measure of the twist of the charge anisotropy, with a finite value signaling the absence of a common vertical mirror plane.

Two particularly simple minimal chiral combinations are
\begin{align}
 (Q_{nn}^{(c)},Q_{n+1,n}^{(s)}),\qquad
 (Q_{nn}^{(s)},Q_{n+1,n}^{(c)}).
 \label{eq:crosspairs}
\end{align}
For equal positive coefficients, these two combinations have opposite signs of $\chi_n$ and hence represent opposite handednesses.
Each combination possesses, in addition to the principal $C_n$ rotation, a set of twofold rotation axes perpendicular to the principal axis.
It therefore realizes a chiral charge distribution with $D_n$ symmetry, up to an appropriate choice of the in-plane coordinate axes.

For the polar group $C_n$, the electric dipole $Q_{10}\propto z$ is also symmetry allowed.
Importantly, this polar component is not required to generate chirality.
Rather, when added to a $D_n$-symmetric cross-pair configuration, it breaks the perpendicular twofold rotations while preserving the principal $C_n$ rotation, thereby reducing the symmetry from $D_n$ to $C_n$.
More generally, $Q_{10}$ is not necessary even for realizing $C_n$ symmetry: a generic four-component combination in Eq.~(\ref{eq:general}) already has $C_n$ symmetry when no perpendicular twofold rotation remains.

\begin{figure}[ht!]
 \centering
 \includegraphics[width=1.0\linewidth]{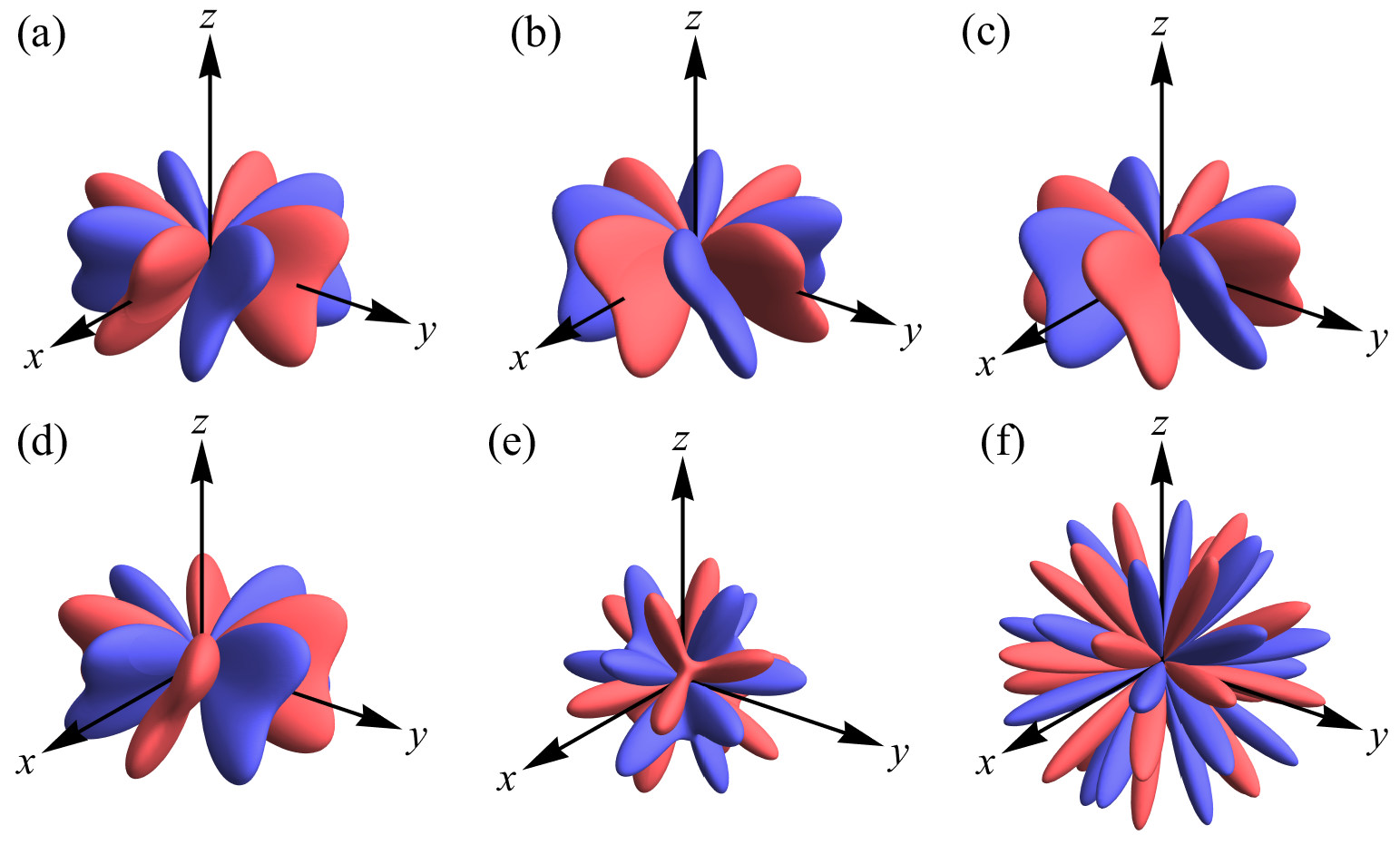}
 \caption{
 Representative chiral atomic charge distributions constructed from electric multipoles.
 (a),(b) Opposite handednesses for the $D_4$-symmetric combinations $Q_{44}^{(c)}+Q_{54}^{(s)}$ and $Q_{44}^{(c)}-Q_{54}^{(s)}$, respectively.
 (c),(d) Corresponding combinations $Q_{44}^{(s)}+Q_{54}^{(c)}$ and $Q_{44}^{(s)}-Q_{54}^{(c)}$, which are additionally allowed under $C_4$ symmetry.
 (e) Tetrahedral chiral distribution represented by the coexistence of $Q_{32}^{(s)}$ and $Q_{6t}$.
 (f) Octahedral chiral distribution represented by the rank-9 harmonic $Q_9^O$.
 Red and blue represent opposite signs of the angular charge-density anisotropy.
 }
 \label{fig:multipoles}
\end{figure}

Figures~\ref{fig:multipoles}(a)--(d) visualize representative chiral charge distributions for the tetragonal chiral point groups.
For $D_4$, Figs.~\ref{fig:multipoles}(a) and (b) show the two opposite handednesses generated by $Q_{44}^{(c)}\pm Q_{54}^{(s)}$.
Their angular dependences are proportional to $ \sin^4\theta ( \cos4\phi\pm\cos\theta\sin4\phi)$ and reversing the relative sign reverses the sense of the latitude-dependent twist.
In $C_4$, the complementary tesseral components $Q_{44}^{(s)}$ and $Q_{54}^{(c)}$ are also symmetry allowed.
Figures~\ref{fig:multipoles}(c) and \ref{fig:multipoles}(d) show the corresponding combinations $Q_{44}^{(s)}\pm Q_{54}^{(c)}$, which exhibit the same reversal of handedness upon changing the relative sign.
Choosing a generic relative phase between the rank-4 and rank-5 sectors removes the perpendicular twofold rotations and yields exact $C_4$ symmetry.

The remaining chiral point groups require separate consideration.
The $C_1$ case differs from the axial series because no rotational symmetry is imposed.
Within a fixed real tesseral basis, a simple low-rank chiral combination is $ (Q_{10},Q_{11}^{(c)},Q_{22}^{(s)}) \propto (z,x,xy)$.
For generic nonzero coefficients, the resulting charge distribution has no improper symmetry operation.
Although the number of nonzero tesseral components depends on the choice of basis, the essential result is that rank-1 and rank-2 multipoles are sufficient to realize chirality in $C_1$.

The cubic chiral groups $T$ and $O$ exhibit qualitatively different multipole structures.
For the tetrahedral group $T$, a representative minimal chiral combination is
$(Q_{32}^{(s)},Q_{6t}) \propto \bigl(xyz,\, x^4(y^2-z^2)+y^4(z^2-x^2)+z^4(x^2-y^2)\bigr)$.
These two multipoles have opposite spatial parity, and their coexistence removes improper symmetries while preserving all rotations of the chiral tetrahedral group $T$, as illustrated in Fig.~\ref{fig:multipoles}(e).
For the proper octahedral group $O$, by contrast, no totally symmetric odd-parity electric multipole exists below rank nine: the totally symmetric sectors at lower ranks occur only at $l=0,4,6,$ and $8$, all of which are even under spatial inversion.
Consequently, no combination of lower-rank electric multipoles can produce an $O$-symmetric chiral charge distribution, and chirality first appears in the rank-9 harmonic $Q_9^O\propto xyz(x^2-y^2)(y^2-z^2)(z^2-x^2)$.
This harmonic preserves all proper octahedral rotations while changing sign under improper spatial operations, and its characteristic high-rank angular structure is shown in Fig.~\ref{fig:multipoles}(f).

\begin{table*}[t]
\caption{
Representative minimal electric-multipole contents generating local charge-density chirality in the 11 crystallographic chiral point groups.
``Polar'' indicates whether the electric dipole $Q_{10}\propto z$ is symmetry allowed.
The polar component is not included in the minimal chiral combination because it is not required for chirality itself.
The Cartesian and angular forms are given up to normalization and radial factors.
}
\label{tab:classification}
\centering
\renewcommand{\arraystretch}{1.22}
\begin{tabular}{ccclll}
\hline\hline
System & PG & Polar & Minimal combinations & Cartesian form & Angular form\\
\hline
Triclinic
& $C_1$ & Yes
& $(Q_{10},Q_{11}^{(c)},Q_{22}^{(s)})$
& $z+x+xy$
& $c_\theta+s_\theta c_\phi+\frac12s_\theta^2s_{2\phi}$\\
\hline
Monoclinic
& $C_2$ & Yes
& $\begin{array}{l}(Q_{22}^{(c)},Q_{32}^{(s)})\\(Q_{22}^{(s)},Q_{32}^{(c)})\end{array}$
& $\begin{array}{l}(x^2-y^2)+2xyz\\2xy+z(x^2-y^2)\end{array}$
& $\begin{array}{l}s_\theta^2(c_{2\phi}+c_\theta s_{2\phi})\\s_\theta^2(s_{2\phi}+c_\theta c_{2\phi})\end{array}$\\
\hline
Orthorhombic
& $D_2$ & No
& $(Q_{22}^{(c)},Q_{32}^{(s)})$
& $(x^2-y^2)+2xyz$
& $s_\theta^2(c_{2\phi}+c_\theta s_{2\phi})$\\
\hline
Tetragonal
& $C_4$ & Yes
& $\begin{array}{l}(Q_{44}^{(c)},Q_{54}^{(s)})\\(Q_{44}^{(s)},Q_{54}^{(c)})\end{array}$
& $\begin{array}{l}x^4-6x^2y^2+y^4+4zxy(x^2-y^2)\\4xy(x^2-y^2)+z(x^4-6x^2y^2+y^4)\end{array}$
& $\begin{array}{l}s_\theta^4(c_{4\phi}+c_\theta s_{4\phi})\\s_\theta^4(s_{4\phi}+c_\theta c_{4\phi})\end{array}$\\
& $D_4$ & No
& $(Q_{44}^{(c)},Q_{54}^{(s)})$
& $x^4-6x^2y^2+y^4+4zxy(x^2-y^2)$
& $s_\theta^4(c_{4\phi}+c_\theta s_{4\phi})$\\
\hline
Trigonal
& $C_3$ & Yes
& $\begin{array}{l}(Q_{33}^{(c)},Q_{43}^{(s)})\\(Q_{33}^{(s)},Q_{43}^{(c)})\end{array}$
& $\begin{array}{l}(x^3-3xy^2)+z(3x^2y-y^3)\\(3x^2y-y^3)+z(x^3-3xy^2)\end{array}$
& $\begin{array}{l}s_\theta^3(c_{3\phi}+c_\theta s_{3\phi})\\s_\theta^3(s_{3\phi}+c_\theta c_{3\phi})\end{array}$\\
& $D_3$ & No
& $(Q_{33}^{(c)},Q_{43}^{(s)})$
& $(x^3-3xy^2)+z(3x^2y-y^3)$
& $s_\theta^3(c_{3\phi}+c_\theta s_{3\phi})$\\
\hline
Hexagonal
& $C_6$ & Yes
& $\begin{array}{l}(Q_{66}^{(c)},Q_{76}^{(s)})\\(Q_{66}^{(s)},Q_{76}^{(c)})\end{array}$
& $\begin{array}{l}x^6-15x^4y^2+15x^2y^4-y^6\\\quad+z(6x^5y-20x^3y^3+6xy^5)\\6x^5y-20x^3y^3+6xy^5\\\quad+z(x^6-15x^4y^2+15x^2y^4-y^6)\end{array}$
& $\begin{array}{l}s_\theta^6(c_{6\phi}+c_\theta s_{6\phi})\\s_\theta^6(s_{6\phi}+c_\theta c_{6\phi})\end{array}$\\
& $D_6$ & No
& $(Q_{66}^{(c)},Q_{76}^{(s)})$
& $\begin{array}{l}x^6-15x^4y^2+15x^2y^4-y^6\\\quad+z(6x^5y-20x^3y^3+6xy^5)\end{array}$
& $s_\theta^6(c_{6\phi}+c_\theta s_{6\phi})$\\
\hline
Cubic
& $T$ & No
& $(Q_{32}^{(s)},Q_{6t})$
& $\begin{array}{l}xyz+x^4(y^2-z^2)+y^4(z^2-x^2)\\\quad+z^4(x^2-y^2)\end{array}$
& $\begin{array}{l}s_\theta^2c_\theta s_{2\phi}-\frac1{16}s_\theta^6c_{6\phi}\\+\left[s_\theta^2c_\theta^4-s_\theta^4c_\theta^2+\frac1{16}s_\theta^6\right]c_{2\phi}\end{array}$\\
& $O$ & No
& $(Q_9^O)$
& $xyz(x^2-y^2)(y^2-z^2)(z^2-x^2)$
& $\begin{array}{l}s_\theta^4c_\theta[-2(1-10c_\theta^2+17c_\theta^4)s_{4\phi}\\\qquad\qquad+s_\theta^4s_{8\phi}]\end{array}$\\
\hline\hline
\end{tabular}
\end{table*}

We summarize the minimal chiral multipole realizations for all 11 crystallographic chiral point groups in Table~\ref{tab:classification}.
For compactness, we use $s_\theta=\sin\theta$, $c_\theta=\cos\theta$, $s_{m\phi}=\sin m\phi$, and $c_{m\phi}=\cos m\phi$.
Representative minimal combinations are listed together with their Cartesian and angular forms.
The ``Polar'' column specifies whether $Q_{10}\propto z$ is symmetry allowed, separately from the multipoles responsible for chirality.

An important consequence of the above classification is that crystallographic chirality need not be encoded in a restricted atomic charge density.
For a single-center product of spinless atomic orbitals with angular momenta $L_1$ and $L_2$, the accessible electric-multipole ranks obey the Gaunt-selection rules $|L_1-L_2|\leq l\leq L_1+L_2$ and $L_1+L_2+l={\rm even}$.  
Therefore, a complete orbital basis truncated at maximum angular momentum $L_{\rm max}$ has the upper bound $l \leq 2L_{\rm max}$ for the scalar charge density.
Thus, a single-center $s$-$p$-$d$-$f$ basis supports multipoles only up to rank six, whereas the minimal chiral constructions for $C_6$ and $D_6$ require rank seven and that for $O$ requires rank
nine.
Consequently, the handedness of these crystallographic environments cannot be encoded in the local scalar charge density within an $s$-$p$-$d$-$f$ basis.

The $C_6$ case provides a simple illustration.
Up to rank six, the $C_6$-symmetric charge density can be written as
\begin{align}
 \rho_{C_6}^{(l\leq6)}
 =
 f(\cos\theta)
 +A\sin^6\theta\cos(6\phi-\delta),
 \label{eq:C6rank6}
\end{align}
where $f(\cos\theta)$ contains the $m=0$ components. 
This form necessarily retains vertical mirror planes and is therefore achiral.
The first component capable of removing them is the rank-7 harmonic $Q_{76}^{(s)}$, demonstrating explicitly why an orbital basis extending beyond $f$ orbitals is required for $C_6$ charge chirality.

This rank criterion also clarifies the contrast between elemental Te/Se and B20-type compounds.  
A Te/Se atomic site has $C_2$ symmetry, but an on-site spinless $s$-$p$ basis reaches only rank two and therefore cannot realize the minimal $C_2$ charge-chiral pair $(l,m)=(2,2)$ and $(3,2)$.  
Its structural chirality can instead be encoded in spinless orbital-exchange hopping between the $p_z$ and $(p_x,p_y)$ orbitals~\cite{oiwa2025predominant}. 
By contrast, the $C_3$ of transition metal sites in B20-type compounds can realize the minimal $C_3$ charge-chiral pair $(l,m)=(3,3)$ and $(4,3)$ within $p$-$d$ and $d$-$d$ hybridizations, consistent with the observed chiral valence-electron-density twist~\cite{aoyagi2026experimental}.

Finally, let us comment on the relation between the present construction and the electric toroidal monopole $G_0$, which represents a time-reversal-even pseudoscalar in the general multipole classification~\cite{kusunose2022generalization, hayami2024unified}.
The $G_0$ multipole has also been used as a microscopic indicator of electronic chirality in chiral structures, including elemental Te~\cite{oiwa2025predominant} and twisted methane~\cite{inda2024quantification}.
Since $G_0$ is not itself a component of the scalar charge density~\cite{hayami2018microscopic}, its pseudoscalar character is instead encoded in the relative configuration of electric multipoles.
For the axial groups $C_n$ and $D_n$ with $n=2,3,4,6$, the signed relative orientation between the rank-$n$ and rank-$(n+1)$ anisotropies provides such a composite pseudoscalar.
In this sense, the present charge-density formulation may be regarded as a real-space electric-multipole representation of a $G_0$-type chiral degree of freedom.

In summary, we established an electric-multipole ($Q_{lm}$) criterion for local atomic charge chirality.
For the axial $C_n$ and $D_n$ groups, chirality is encoded in the relative orientation of same-$m$ anisotropies of consecutive ranks $l = n$ and $n+1$, while the cubic groups $T$ and $O$ exhibit distinct multipole realizations.
The resulting rank criterion determines whether crystallographic handedness can be encoded within a given atomic orbital basis, thereby linking orbital content to local charge chirality and providing a practical criterion applicable to experimentally measured and first-principles charge densities~\cite{oiwa2025symmetry,muller2026multipoles}.

The authors thank N. Kanazawa, S. Aoyagi, and S. Kitou for fruitful discussions.
This research was supported by JSPS KAKENHI Grant Numbers
JP22H00101, JP23H04869, and JP26K22277, and by JST CREST (JPMJCR23O4 and JPMJCR2641) and JST FOREST (JPMJFR2366).

\bibliography{ref.bib}
\bibliographystyle{jpsj}

\end{document}